\documentclass[conference]{IEEEtran}
\usepackage{cite}
\usepackage{subcaption}
\usepackage{amsmath,amssymb,amsfonts}
\usepackage{algorithmic}
\usepackage{graphicx}
\usepackage{comment}
\usepackage{textcomp}
\usepackage[table]{xcolor}
\usepackage{xcolor}
\usepackage{multirow}
\usepackage{hyperref}
\usepackage{url}
\usepackage{soul}
\newcommand{\furl}[1]{\footnote{\scriptsize \url{#1}}}
\colorlet{punct}{red!60!black}
\definecolor{background}{HTML}{EEEEEE}
\definecolor{delim}{RGB}{20,105,176}
\colorlet{numb}{magenta!60!black}
\usepackage[hang,flushmargin]{footmisc}
\def\BibTeX{{\rm B\kern-.05em{\sc i\kern-.025em b}\kern-.08em
    T\kern-.1667em\lower.7ex\hbox{E}\kern-.125emX}}
\begin{document}

\title{OER Recommendations to Support Career Development}

\author{\IEEEauthorblockN{Mohammadreza Tavakoli}
\IEEEauthorblockA{\textit{Leibniz Information Centre} \\
\textit{for Science and Technology} \\
Hannover, Germany \\
reza.tavakoli@tib.eu}
\and
\IEEEauthorblockN{Ali Faraji}
\IEEEauthorblockA{\textit{RWTH Aachen University} \\
Aachen, Germany \\
abdolali.faraji@rwth-aachen.de}
\and
\IEEEauthorblockN{Stefan T. Mol}
\IEEEauthorblockA{\textit{University of Amsterdam} \\
Amsterdam, Netherlands \\
s.t.mol@uva.nl}
\and
\IEEEauthorblockN{G\'abor Kismih\'ok}
\IEEEauthorblockA{\textit{Leibniz Information Centre} \\
\textit{for Science and Technology} \\
Hannover, Germany \\
gabor.kismihok@tib.eu}
}

\maketitle

\begin{abstract}
This \emph{Work in Progress} \emph{Research} paper departs from the recent, turbulent changes in global societies, forcing many citizens to re-skill themselves to (re)gain employment. Learners therefore need to be equipped with skills to be autonomous and strategic about their own skill development. Subsequently, high-quality, on-line, personalized educational content and services are also essential to serve this high demand for learning content. Open Educational Resources (OERs) have high potential to contribute to the mitigation of these problems, as they are available in a wide range of learning and occupational contexts globally. However, their applicability has been limited, due to low metadata quality and complex quality control. These issues resulted in a lack of personalised OER functions, like recommendation and search. Therefore, we suggest a novel, personalised OER recommendation method to match skill development targets with open learning content. This is done by: 1) using an OER quality prediction model based on metadata, OER properties, and content; 2) supporting learners to set individual skill targets based on actual labour market information, and 3) building a personalized OER recommender to help learners to master their skill targets. Accordingly, we built a prototype focusing on \emph{Data Science} related jobs, and evaluated this prototype with 23 data scientists in different expertise levels. Pilot participants used our prototype for at least 30 minutes and commented on each of the recommended OERs. As a result, more than 400 recommendations were generated and \emph{80.9\%} of the recommendations were reported as useful.
\end{abstract}


%context, source, integration, UI

\section{Introduction}
In recent years, we have been facing with a quickly evolving labour market leading to several educational challenges caused by the gap between what educational institutions offer and what is actually required by labour market \cite{colombo2018applying,wowczko2015skills,mcgill2009defining}. Furthermore, with the emergence of COVID-19 pandemic, and the collapse of a number of local and global industries, very large amount of employees need to re-skill themselves in order to regain employment. Therefore, updating individuals' skills and knowledge, to adapt to post COVID-19 labour markets, is a deep concern of our current societies, including educational decision makers, employers, job holders/seekers alike. To tackle this enormous challenge, an individual level self development plan based on dynamic labour market information and Open Educational Resources (OERs) could be critically important~\cite{tavakoli2020labour}. The advantage of OERs is that they are provided by large number of experts in very different contexts (e.g. country, city, expertise level, language), which is important for labour market driven education. 

Previous efforts to build educational search and recommender systems using the rapidly growing amount of OERs \cite{wan2018learning,sun2017towards,duffin2007oer} revealed the lack of high-quality OER metadata, and quality control~\cite{tavakoli2020quality}. These issues seriously limit the deployment of high-quality OER search and recommendation services~\cite{sun2018heuristic,chicaiza2017recommendation}. 

In this paper, therefore, we showcase our attempt to extend our personalised, labour market information based OER recommender~\cite{tavakoli2020labour} with OER quality prediction models~\cite{tavakoli2020recommender} and enhanced metadata scoring~\cite{tavakoli2020quality}. Furthermore, we show the integration process of 7 different OER repositories, and our latest learner dashboard prototype focused on \emph{Data Science} related jobs. Finally, we describe our efforts to validate our prototype by 23 data science subject matter experts.

\section{State of the Art}
We can group the related studies into two main categories: A) OER quality assessment, and B) OER recommender systems.  
\subsection{OER Quality Assessment}
Due to the large amount of OERs being created and uploaded by people around the world, manual quality control of OERs is getting more and more complicated, if not impossible. In general, high-quality metadata is a mandatory property for providing data driven services~\cite{kiraly2018measuring}. In the area of OERs, low-quality metadata not only reduces the discoverability of OERs, but also has a strong negative effect on their usability~\cite{ushakova_2015}. Accordingly, some studies attempted to define metrics (e.g. completeness and consistency of metadata~\cite{pelaez2017metadata}, provenance, and accuracy of metadata~\cite{romero2018exploring}) in order to assess the quality of OER metadata. To support this, \cite{tavakoli2020quality} showed that a metadata scoring method can be used to define a model to predict OER quality.

Additionally,~\cite{romero2019proposal} provides a quality assessment framework based on existing developmental models in the area of e-learning, semantic-based methods, and NLP techniques. Moreover, in the case of open educational videos, quality can be predicted on the basis of video transcripts, popularity metrics (i.e. rate, likes, dislikes), and length~\cite{tavakoli2020recommender}. 

\subsection{OER Recommendation Systems}
The literature on high-quality OER recommender systems is very limited \cite{chicaiza2017recommendation}. However, in recent years, some studies built recommendation algorithms based on ontologies, linked data, and open source RDF data to leverage semantic content~\cite{chicaiza2017recommendation,sun2017towards}. As an example, \cite{wan2018learning} defined an ontology for learners, learning objects, and their environments in order to provide adaptive recommendations, based on similarities between object properties. \cite{sun2018heuristic} examined the Cold Start problem \cite{lam2008addressing} in the case of new micro OERs, by defining rules based on recommended sequences of learning objects using existing ontologies. Furthermore, ~\cite{tavakoli2020labour} built an OER recommendation system for learners in order to help them to achieve skill based learning objectives.

\section{Data Collection}
In this section, we explain the steps we followed to collect relevant and reliable data sources for our recommender prototype, focusing on data science related skills and jobs.
\subsection{Labour Market Information}
We applied the method proposed by ~\cite{tavakoli2020labour} on 500 job vacancies, published in February and January 2020, which resulted in the extraction of the 16 most relevant and required skills for data science jobs. Additionally, we used Wikipedia python API\footnote{\url{https://pypi.org/project/wikipedia/}} and crawled  Wikipedia content describing these 16 extracted skills.
%\footnote{Complete list of skills with their properties: \url{http://s000.tinyupload.com/?file_id=87025321518907345755}}
\subsection{OER Resources}
Regarding the educational content, we collected more than 700 OERs from the following repositories: Youtube, MIT OpenCourseWare\footnote{\url{https://ocw.mit.edu/}}, Skills Commons\footnote{\url{https://www.skillscommons.org/}}, OER Commons\footnote{\url{https://www.oercommons.org/}}, Wisc-Online\footnote{\url{https://www.wisc-online.com/}}, Khan Academy\footnote{\url{https://www.khanacademy.org/}}, and Wikipedia by using their APIs and/or search services. For each OER, we store the following fields (if available), in order to provide effective recommendations: source, title, description, target skill, level (Beginner, Intermediate, Advanced, Master), URL, length, transcription, view count, rating (or combination likes and dislikes), ranking position score (reverse of ranking position in the source repository search: $1/ranking\_position\_in\_repository\_search$), and transcription similarity with the target skill description (calculated by the cosine similarity between the transcription vector and the target skill description vector using pre-trained 300-dimensional Glove vectors \cite{pennington2014glove}).

\section{Methods}
In this section, we detail the most important components of our OER recommender prototype. These components are responsible for OER quality control, recommendation generation, and the interaction with the learner. 

\subsection{OER Quality Predictor}
The aim of this component is to predict the overall quality of the collected OERs from different repositories, and define an automatic quality control mechanism. 
\subsubsection{Quality Prediction Based on Properties}
As a first step, we established a measure for OER-skill matching. For this measure we used procedures from \cite{tavakoli2020recommender}\footnote{With \emph{F1-score} of \emph{86.3\%}; The data-sets and the implementation protocols are included in the paper.} to define a random forest model that classifies OERs into two groups: \emph{Fit for Achieving a Skill}, and \emph{Not Fit for Achieving a Skill}. The model uses the following OER properties to execute this prediction: length, rate, transcription similarity with the target skill description, view count, and ranking position score.

Furthermore, we had to find a solution for missing values in our prediction model, since not all OERs contained all required features. For this problem, we built a number of alternative random forest models with various combinations of the above-mentioned features, and selected the one with the highest F1-score. It should be mentioned that we applied alternative models on the dataset defined by \cite{tavakoli2020recommender}, and got a minimum accuracy of \emph{77.2 \%}. The feature-set was adjusted towards maximum accuracy, but in some cases we had to rely on models with fewer features (from all desired OER features) than optimal, which resulted in a slightly reduced accuracy.

\subsubsection{Quality Prediction Based on Metadata}
Previously we proposed a metadata scoring model based on 8,887 crawled OERs from \emph{Skill Commons}~\cite{tavakoli2020quality}\footnote{The data-sets and the implementation protocols are included in the paper.}. This model divides OERs in two groups: \emph{OERs with Manual Quality Control}, and \emph{OERs Without Quality Control}. At the same time, we also showed that there is a tight relationship between OER metadata quality and OER quality control processes. This means, that the more an OER passes quality controls, the higher the probability of containing high-quality metadata is. However, due to the rapidly growing amount of OERs, manual quality control is getting more and more impossible ~\cite{tavakoli2020quality}. As a consequence, we built a model that predicts OER quality based on their metadata with \emph{F1-score} of \emph{94.6\%}. Therefore, in this paper we also capitalized on this scoring and prediction model, to qualify metadata, and also to predict the quality of OERs we collected form 7 distinct repositories.  

\subsubsection{Quality Control Based on the Defined Models}
To apply an automatic quality control process on the collected OERs, we used the above-mentioned models in a probability mode, which shows the probability of the target class in the classifiers. Therefore, for each OER, based on their properties and metadata, we applied both of our prediction models and defined two float numbers between 0 and 1 to determine the OER quality scores. Finally, we removed OERs with quality scores less than 0.5 (altogether 184 OERs) from the list of potential OERs for recommendation.

\begin{figure*}
    \centering
    \includegraphics[width=\textwidth]{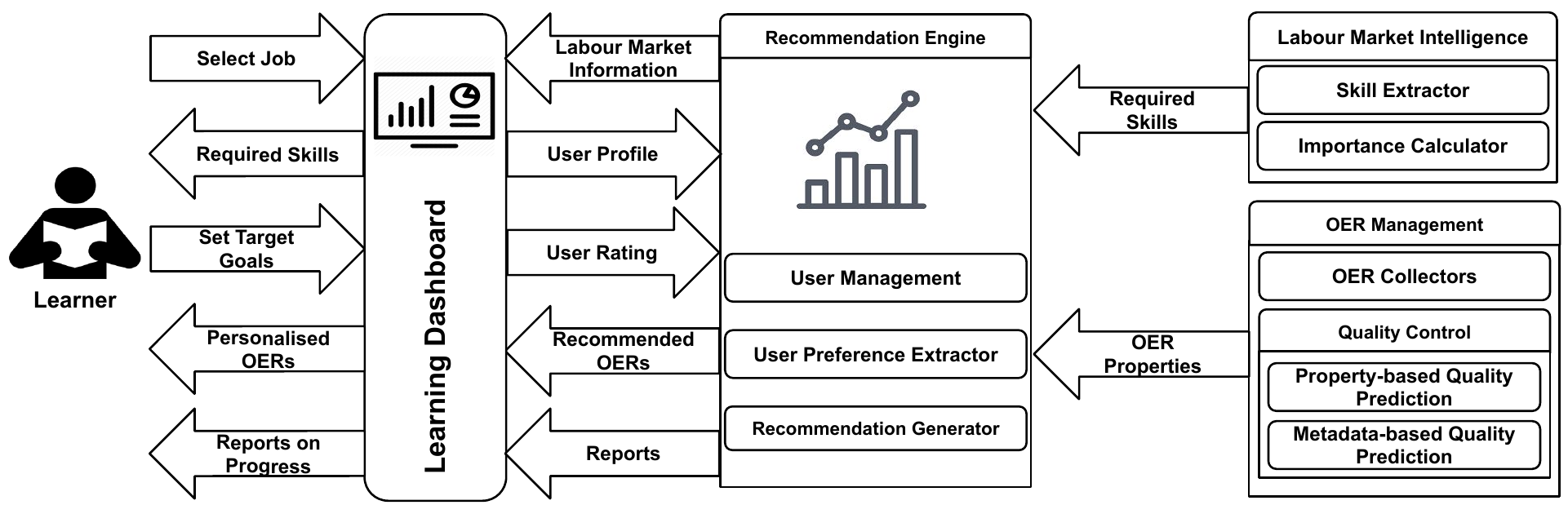}
    \caption{Components of our Labour Market Driven Personalised OER Recommender System}
    \label{fig:structure}
\end{figure*}

\subsection{Recommendation Generation}
In order to build our OER recommender system, we created a 6-dimensional vector of $X$ for each OER including:
\begin{itemize}
    \item Normalised length    
    \item Normalised rate
    \item Text similarity with target skill description
    \item Metadata based quality probability
    \item Property based quality probability
    \item Source (containing 7 binary values to identify the 7 OER repositories we used for this study)
\end{itemize}
Respectively, for each learner, we define a 6-dimensional vector $P$ as a preference vector that contains a weight (between 0 and 1) for each parameter in $X$.

Moreover, to provide personalised recommendations for learners, we rely on the following features describing learners' context:
\begin{itemize}
    \item Country
    \item City
    \item Job experience including skills, expertise levels in each skill (Beginner, Intermediate, Advanced, Master) and industry\footnote{We defined industries based on \url{https://en.wikipedia.org/wiki/Statistical_Classification_of_Economic_Activities_in_the_European_Community}}
    \item Birth date    
    \item Gender (optional)
\end{itemize}

The goal is to find the best weights (\emph{P} vector) for each learner. Therefore, at learner registration, we initialise the $P$ vector for each learner, based on preference vectors of similar learners in the learner's context (e.g. Country, City, Job experience). Subsequently, we use Gradient Descend to optimize the preference vector based on users' ratings by minimizing the following loss function:
\begin{equation} \label{eq:loss}
  Loss Function = \sum_{o=recommended\_OERs} |(P * X_o) - Y_o|
\end{equation}
where $X_o$ is the 6-dimensional vector of an OER \emph{o} and $Y_o$ is the satisfaction rating of the learner for that particular OER \emph{o}.

Finally, to recommend an OER to a learner \emph{u} for a particular target skill \emph{s}, the system retrieves all OERs with the level that learner \emph{u} has in skill \emph{s}, and uses cosine similarity to recommend the OER with the closest $X$ vector to the user preference vector ($P$).

\subsection{Learning Dashboard}
In this section, we explain the features of the implemented learner dashboard prototype, focusing on data science related jobs, using the above-mentioned components\footnote{You can find a demo of the prototype from: \url{https://github.com/rezatavakoli/FIE2020}}. As an illustration, Figure~\ref{fig:structure} shows the different components of our system and their interaction.

When a learner uses our recommender for the first time, (s)he has to go through a two step registration process, which is critical to initialise essential learner features for effective recommendations:

\subsubsection{Profile Creation}
Learners register in our system by adding their demographic features (e.g. country, city, and birth date). Subsequently, they can also add their previous job experience(s) to their profile. Based on their previous jobs, our system recommends related skills (based on our labour market analysis) for their profile. Learners also need to set their expertise level(s) for each skill. 

\begin{table*}
\caption{Users Satisfaction Through Evaluation}\label{tab:tbl-eval}
\resizebox{\textwidth}{!}{
\begin{tabular}{|c|c|c|c|c|c|c|c|}
\hline
\rowcolor[HTML]{C0C0C0} 
\textbf{\shortstack{Progress in Evaluation (\%)}} & \textbf{Very Dissatisfied (\%)} & \textbf{Dissatisfied (\%)} & \textbf{OK (\%)} & \textbf{Satisfied (\%)} & \textbf{Very Satisfied (\%)} & \textbf{Report as Useful (\%)} \\ \hline
0-25 & 5 & 21 & 37 & 22 & 15 & 74 \\  \hline
25-50 & 4 & 18 & 34 & 27 & 17 & 78 \\  \hline
50-75 & 3 & 14 & 32 & 33 & 18 & 83 \\  \hline
75-100 & 3 & 9 & 33 & 35 & 20 & 88 \\  \hline
\textbf{Total (0-100)} & \textbf{$\approx$4} & \textbf{$\approx$15} & \textbf{$\approx$34} & \textbf{$\approx$29} & \textbf{$\approx$18} & \textbf{$\approx$81 (=80.9)}\\  \hline
\end{tabular}
}
\end{table*}

\subsubsection{Defining Goals}
Once a learner profile is created, our system asks learners to set their desired job. Based on this target occupation, our system shows all skills, which are required for that job (in the desired geographical location of the learner: city, country). Learners can select their skill goals from a list, but also search and add additional skill targets to their profiles\footnote{Screenshot of the goal definition page: \url{https://github.com/rezatavakoli/FIE2020}}.  

\subsection{Educational Content Page}
For each selected skill objective, our system recommends the most relevant OERs. Learners can click and navigate to the physical location of the OER, and proceed with the learning. After learning, users can rate their satisfaction with each OER (from 1: very dissatisfied to 5: very satisfied), request another OER on the same level of difficulty, or ask for OERs with a higher level of difficulty. Moreover, if dissatisfied with the recommendation, learners can either replace the recommended OERs with another recommendation, or mark the OER as irrelevant for their skill target and ask for a replacement OER\footnote{Screenshot of the educational content page: \url{https://github.com/rezatavakoli/FIE2020}}.

\subsubsection{Progress Page and Reports}
At each stage, learners can monitor their progress towards their goals (target job and skills), and consult reports on finished OERs at each expertise level they passed. Moreover, they can retrieve reports on periodically (e.g. monthly) completed OERs, and also the status of their current recommendations (e.g. in progress, finished, changed)\footnote{Screenshot of the progress page: \url{https://github.com/rezatavakoli/FIE2020}}.

\section{Validation}
To validate our approach and the proposed prototype, we invited 23 experts (including 16 PhD students and 7 university instructors) in the area of \emph{Data Science} (with minimum 1 year of related teaching experience and 3 years of related industrial experience)  and asked them to work with our prototype for at least 30 minutes. It should be mentioned that participants had different expertise levels regarding the required skills for a data scientist job as they had various range of teaching and industrial experience. Phd students had related teaching experience between 1 year and 6 years and related industrial experience between 3 years to 10 years. These numbers for instructors were between 7 to 16 years for teaching experience and 8 to 19 years for industrial experience. Accordingly, when experts registered in our system, 22\% of skill expertise levels were set to Beginner, 26\% were set to Intermediate, 31\% were set to Advanced, and 21\% were set to Master.

After creating their profile and setting their goals, participants started working with their learning dashboard, including their recommended OERs. Each participant generated at least 17, but maximum 20 recommendations. During the validation, altogether 415 OERs were recommended, and 336 of them (\emph{80.9\%}) were reported as useful (\emph{OK}, \emph{Satisfied}, or \emph{Very Satisfied}). To illustrate how the quality of recommendations increased over time, we grouped all recommendations into four clusters. Each quarter contains a bit more than 100 recommendations, and Table~\ref{tab:tbl-eval} shows participants' satisfaction rate for each cluster (over time).  The \emph{Report as Useful} column reveals that the more the users work with our system, the more satisfactory recommendations they receive, since our system attempts to identify learner preferences and provide personalised recommendations. Moreover, based on our quality prediction and quality control process, more than 70\% of the recommendations satisfied learners right from the beginning. As a consequence, we believe that this exercise revealed that our approach and the implemented prototype can potentially provide high-quality support to learners working towards labour market's demanded skills. 

\section{Conclusion and Future Work}
This study is one of the first steps to 1) empower learners to take control and responsibility for their own skill development, 2) improve skills on the basis of labour market information and personalised OER recommendation, and 3) progress the literature on OER quality control. As a long term vision, we expect that learners will show enhanced self-regulation, and spend less effort to find relevant, high-quality OERs. This approach also contributes to OER property identification accuracy, which is essential to increase OER (re)usability.

In this paper, we also demonstrated an OER recommender system prototype, which collects OERs from 7 different OER repositories, predicts their quality, and applies automatic quality control. Learners interact with the recommender through a dashboard, in which they can search for their desired job, display the list of required skills, set their level of expertise for each skill, set their learning context, and receive relevant OER recommendations accordingly. During their learning process, learners rate their satisfaction with recommendations, and update their learning preference vector. This strategy detects changes in learner profiles and fine tune the precision of recommendations.

For the next steps, we plan to add more OER repositories, extract more learner and OER properties, and use (quasi-)experimental designs for further developing and validating our prototype in a number of use cases.

\bibliographystyle{IEEEtran}
\bibliography{fie}

\end{document}